\documentclass[openacc]{rstransa}
\usepackage{tikz,tkz-euclide}
\usetikzlibrary{shapes}
\usetikzlibrary{positioning}
\usepackage{adjustbox}

\begin{document}

\title{Exploring the origin of Turbulent Taylor rolls}

\author{
V. Jeganathan$^{1}$, K. Alba$^{2}$ and R. Ostilla-M\'onico$^{1,3}$}

\address{$^{1}$Department of Mechanical Engineering, University of Houston, Houston 77004, USA\\
$^{2}$Department of Engineering Technology, University of Houston, Houston 77004, USA\\
$^{3}$Escuela Superior de Ingener\'ia, Universidad de C\'adiz, C\'adiz, Spain }

\subject{xxxxx, xxxxx, xxxx}


\corres{R. Ostilla-M\'onico\\
\email{rodolfo.ostilla@uca.es}}

\begin{abstract}
Since Taylor's seminal paper, the existence of large-scale quasi-axisymmetric structures has been a matter of interest when studying Taylor-Couette flow. 
In this manuscript, we probe their formation in the highly turbulent regime by conducting a series of numerical simulations at a fixed Reynolds number $Re_s=3.6\times 10^4$ while varying the Coriolis parameter to analyze the flow characteristics as the structures arise and dissipate. We show how the Coriolis force induces a one-way coupling between the radial and azimuthal velocity fields inside the boundary layer, but in the bulk there is a two-way coupling that causes competing effects. We discuss how this complicates the analogy of narrow-gap Taylor-Couette to other convective flows. We then compare these statistics to a similar shear flow without no-slip boundary layers, showing how this double coupling causes very different effects. We finish by reflecting on the possible origins of turbulent Taylor rolls.
\end{abstract}




\maketitle

\section{Introduction}

In 1923, G.~I.~Taylor published his celebrated paper ``Stability of a Viscous Liquid contained between Two Rotating  Cylinders'', where he derived the basic conditions for instability of what would be later called Taylor-Couette flow (TCF), i.e.~the flow between two coaxial and independently rotating cylinders \cite{taylor1923viii}. In this paper, he examined the linear stability of the base flow to axisymmetric perturbations, calculating the critical Reynolds number at which these perturbations would become unstable. He also calculated the streamlines of the resulting flow field which had a peculiar structure that closely matched what was seen in experiments, and filled the entire gap for conditions of stationary outer cylinder and co-rotating cylinders. These flow structures would henceforth be known as Taylor rolls, after him.

After the onset of linear instability, the flow between the cylinders consists only of laminar Taylor rolls. As the Reynolds number is increased, the rolls go through a series of transitions. First, they lose their axisymmetry, and waves appear in the azimuthal direction. This is known as wavy Taylor vortex flow \cite{jones1985transition}. Increasing the Reynolds number leads to a second transition: the onset of time-dependence, resulting in modulated wavy Taylor vortex flow \cite{barenghi1989modulated}. Further increases of the Reynolds number lead to the forming of turbulent fluctuations on top of the rolls, a flow regime known as turbulent Taylor rolls \cite{andereck1986flow}. Here, even if the flow is turbulent, the underlying structure of the rolls remains relatively unaffected, i.e.~ they remain quasi-axisymmetric and axially pinned. 

As the Reynolds number is increased beyond $Re_s\sim\mathcal{O}(10^4-10^5)$ \textcolor{blue}{($Re_s$ is defined below)}, a new regime appears, where depending on the exact combination of differential rotation, the large-scale rolls may vanish (or remain) up to the largest shear Reynolds numbers reached in experiments ($\mathcal{O}(10^6)$) \cite{lathrop1992transition,huisman2014multiple}. While this transition is sometimes identified with the transition to the ``ultimate'' regime where both boundary layer and bulk become turbulent, the identification is not very straight forward due to the presence and disappearance of quiescent regions in the boundary layers \cite{ezeta2020double}. Setting aside this issue, what is clear is that in this regime both experiments and simulations show that the rolls appear in a smaller region of parameter space, which often does not coincide with the largest centrifugal instability. Sacco \emph{et al.} showed that for narrow-gap TCF at $Re_s=3.6\times 10^4$, large-scale structures which are axisymmetric and axially pinned persist for certain combinations of mild anti-cyclonic rotation and shear, and that they are preserved at the limit of vanishing curvature as TCF becomes rotating Plane Couette flow (RPCF) \cite{sacco2019dynamics}. Furthermore, they showed that as the Reynolds number increased, the rolls tended to ``empty'' themselves of vorticity, which on average migrated to the boundaries even if the roll shape appears to be largely the same. This resulted in a different flow behaviour than the original Taylor rolls.

\begin{figure}

\begin{center}
\begin{tikzpicture}[scale=0.45]

\draw[-][black,ultra thick, domain=-40:40] plot ({4+4*sin(\x)}, {4*cos(\x)-3});
\draw[-][black,ultra thick,domain=-40:40] plot ({4+9*sin(\x)}, {9*cos(\x)-3});

\draw[<->, ultra thin][gray](6.4,0.4) to (9.4,4.1);

\fill[black](7.9,2.2) node [scale=1,anchor=west]{$\hat d$};

\draw[<-][blue,domain=-8:0] plot ({4+9.5*sin(\x)}, {9.5*cos(\x)-3});
\draw[->][red,domain=0:16] plot ({4+3.5*sin(\x)}, {3.5*cos(\x)-3});

\fill[black](4,6.5) node [scale=1,anchor=west]{$\hat{\omega}_o$};
\fill[black](4,0.5) node [scale=1,anchor=east]{$\hat{\omega}_i$};

\draw (0,-1.5) node {};

\end{tikzpicture}
\hspace{1.4cm}
%
\begin{tikzpicture}[scale=0.45]
\draw[blue,ultra thick](0,5)-- (10,5);
\draw[red,ultra thick](0,0)-- (10,0);


\fill[black](5,5.5) node [scale=1,anchor=center]{$\hat{T}$};
\fill[black](5.,-0.5) node [scale=1,anchor=center]{$\hat T + \hat \Delta $};
\draw[<->, ultra thin][gray](9.6,0.1) to +(0,4.8);
\draw[->, thin](8,3.9) to (8,3.1);
\fill[black](8,3.5) node [scale=1,anchor=east]{$\hat g$};
\fill[black](9.6,2.5) node [scale=1,anchor=west]{$\hat H$};

\draw (0,-1.5) node {};

\end{tikzpicture}

\begin{tikzpicture}[scale=0.45]
\draw[black,ultra thick](0,5)-- (10,5);
\draw[black,ultra thick](0,0)-- (10,0);

\draw[->][black,ultra thin](5,5.5)--+ (1.3,0);
\draw[->][black,ultra thin](5,-0.5)--+ (-1.3,0);

\draw[->][black,domain=0:270] plot ({5+0.5*cos(\x)}, {2.5+0.5*sin(\x)});
\filldraw[black] (5,2.5) circle (2pt) node[anchor=east] {};


\draw[<->, ultra thin][gray](9.6,0.1) to +(0,4.8);
\fill[black](9.6,2.5) node [scale=1,anchor=west]{$\hat d$};

		
\fill[black](3.25,5.5) node [scale=1,anchor=west]{$\hat{U}/2$};
\fill[black](5,-0.5) node [scale=1,anchor=west]{$\hat{U}/2$};

\fill[black](4.5,2.5) node [scale=1,anchor=east]{$\hat{\Omega}_{rf}$};

\end{tikzpicture}
\hspace{2cm}
\begin{tikzpicture}[scale=0.45]
\draw[black,ultra thick](0,5)-- (10,5);
\draw[black,ultra thick](0,0)-- (10,0);

\draw[black,dash dot](5,0)-- +(0,5);

\draw[rotate=90,shift={(0,-5)}] (0,1.5) cos (2.5,0) sin(5,-1.5);curve

\draw[->][black,ultra thin](5,0.6)--+ (-1.3,0);
\draw[->][black,ultra thin](5,1.3)--+ (-1,0);
\draw[->][black,ultra thin](5,3.7)--+ (1,0);
\draw[->][black,ultra thin](5,4.4)--+ (1.3,0);

\draw[->][black,domain=0:270] plot ({5+0.5*cos(\x)}, {2.5+0.5*sin(\x)});
\filldraw[black] (5,2.5) circle (2pt) node[anchor=east] {};

	
\fill[black](4,5.5) node [scale=1,anchor=west]{$\hat{\tau}=0$};
\fill[black](4,-0.5) node [scale=1,anchor=west]{$\hat{\tau}=0$};

\draw[<->, ultra thin][gray](9.6,0.1) to +(0,4.8);
\fill[black](9.6,2.5) node [scale=1,anchor=west]{$\hat d$};

\fill[black](4.5,2.5) node [scale=1,anchor=east]{$\hat{\Omega}_{rf}$};

\draw (0,-0.85) node {};
\end{tikzpicture}
\end{center}
\caption{Two-dimensional schematics the canonical flows discussed. Top: Taylor-Couette flow (left) and Rayleigh-B\'enard convection (right). Both systems are conventionally thought of as convective flows, with TCF being driven by the centrifugal forces arising from the differential rotation of the cylinders, while RBC is driven by the buoyancy forces due to the temperature difference $\hat \Delta$ between top and bottom plates.
Bottom: Plane Couette flow (left), where the shear forcing takes place through different plate velocities and Waleffe flow, where the shear forcing takes place via sinusoidal body forcing (right).
The third dimension is omitted for clarity.}
     \label{fi:sch}
\end{figure}
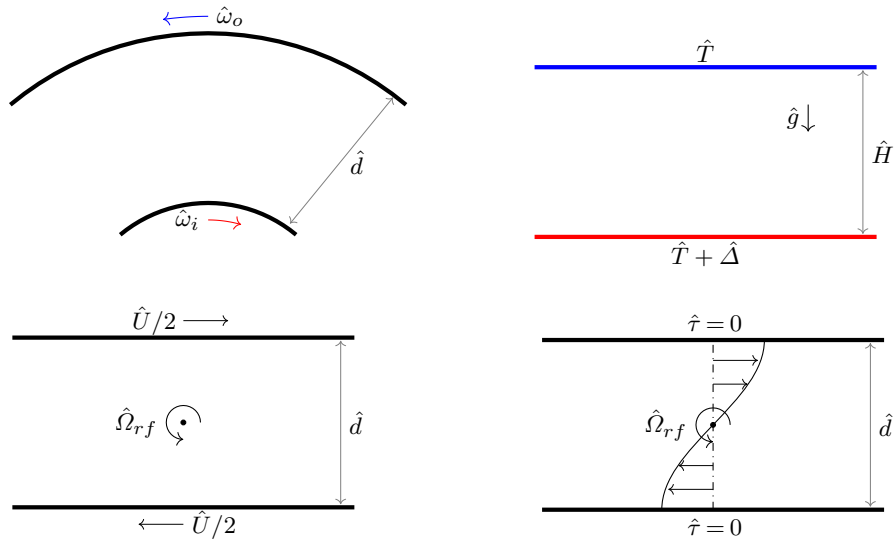

These puzzling observations raise questions on the centrifugal origin of the rolls for high Reynolds number. We can contrast Taylor rolls to the large-scale structures present in another convective flow, i.e.~Rayleigh-B\'enard convection (RBC), the flow in a fluid layer heated from below and cooled from the top \textcolor{blue}{(c.f.~Fig.~\ref{fi:sch} for a schematic of both flows)}. This is hardly a novel concept, RBC and TCF are commonly referred to as ``twins'' of turbulence research, as they possess many similarities and have been productively studied together \cite{busse2012twins,eckhardt2007torque}. In a recent manuscript, Eckhardt \emph{et al.} \cite{eckhardt2020exact} showed an exact correspondence between two-dimensional RBC and two-dimensional three-component Rotating Plane Couette flow (2D3C RPCF, which can be seen as axisymmetric TCF in the limit of vanishing curvature), where the streamwise (azimuthal) velocity played the role of the temperature. Because two-dimensional RBC shows the presence of large-scale structures even at Rayleigh numbers of $Ra\sim\mathcal{O}(10^{14})$, this would roughly correspond to large scale structures persisting for 2D3C RPCF at $Re_s\approx10^7$. This hints that the elusive appearance of the rolls in parameter space is an inherently three-dimensional phenomena of RPCF/TCF. Furthermore, despite the closeness of TCF and RBC, the vanishing of large scale structures is something particular to TCF and not to RBC, as RBC does not show analogous phenomena if the Prandtl number is kept to unity and the Rayleigh number is increased \cite{stevens2018turbulent,pandey2018turbulent,krug2020coherence}, further complicating the analogy between Taylor rolls and convective structures.

In this manuscript, we attempt to explore a possible reason for the different behaviour of the structures in 3D RPCF/TCF and RBC: in RBC, the temperature couples to the vertical velocity as a body force, but in TCF and RPCF, aside from the complex effects of shear, the solid body rotation induces an additional force between the azimuthal/streamwise and radial/wall-normal velocities that is two-way. The radial velocity affects the azimuthal velocity, and viceversa.  By changing the strength of the Coriolis forces, we explore this two-way forcing in diverse regions of the flow, and attempt to rationalize the formation of Taylor rolls in low-curvature TCF. 

To develop these ideas, we will also compare the TCF statistics to results obtained from a similar statistical analysis involving rotating Waleffe flow (RWF), i.e.~the shear flow between two stress-free plates which is forced by a sinusoidal body force, and can be seen as a boundary layer-less RPCF \cite{waleffe1997self}, \textcolor{blue}{ c.f.~Fig.\ref{fi:sch} for comparison between RPCF and RWF}. Recently, Farooq \emph{et al.}~\cite{farooq2020large} showed that streamwise-invariant and spanwise-fixed structures similar to Taylor rolls were present for high-$Re$ RWF when there was a mild anti-cyclonic rotation, very similar to TCF. By comparing the two-way coupling in TCF and RWF, we can further understand the origins of fixed large-scale structures in shear flows.

The manuscript is organized as follows. In the next section we present the numerical details. We then follow by discussing the effect of Taylor rolls on the global statistics of TCF. We quantify the effect of the two-way coupling on local statistics, showing that solid body rotation imposes additional correlations in the boundary layer which are otherwise not present in the bulk. We follow by comparing this to RWF, where the effect is the opposite: rotation imposes further correlations in the bulk and not the boundary layer. We finish with a reflection on the origin of the Taylor rolls and give an outline for further research.

\section{Numerical Details}

\textcolor{blue}{For notation purposes, from here we use hatted symbols to denote dimensional variables ($\hat{\phi}$), and non-hatted symbols ($\phi$) to denote dimensionless variables.}

We perform direct numerical simulations of Taylor-Couette flow in a rotating reference frame by solving the incompressible \textcolor{blue}{non-dimensional} Navier-Stokes equations:

\begin{align}\label{1.1}
\frac{\partial\textbf{u}}{\partial t} + \textbf{u}\cdot\nabla\textbf{u} + R_\Omega (\textbf{e}_z \times \textbf{u})= -\nabla p + Re_s^{-1} \nabla^2\textbf{u}
\end{align}

\noindent with the incompressibility condition 

\begin{align}\label{1.2}
\nabla\cdot\textbf{u}=0,
\end{align}

\noindent where \textbf{u}=$(u_r,u_\theta,u_z)$ is the non-dimensional velocity, $t$ is the non-dimensional time, $Re_s$ the shear Reynolds number defined below, $\textbf{e}_z$ the unit vector in the axial direction, $p$ the non-dimensional pressure and $R_\Omega$ the Coriolis parameter defined below. 

The rotating frame is chosen such that the velocities of both cylinders are equal and opposite, $\pm \hat{U}/2$ in dimensional terms. The equations are non-dimensionalized using this velocity $U$ and the dimensional gap width $\hat{d}$. This results in two non-dimensional control parameters, the shear Reynolds number $Re_s=\hat{U}\hat{d}/\hat{\nu}$ and the Coriolis parameter $R_\Omega=2\hat{\Omega}\hat{d}/\hat{U}$, where $\hat{\nu}$ is the kinematic viscosity of the fluid and $\hat{\Omega}$ is the dimensional rotational velocity of the rotating frame. 

{ \color{blue} We note that $R_\Omega=0$ is equivalent to cylinders rotating with equal and opposite velocities, and modifying this parameter is sufficient to capture any combination of cylinder rotation except solid body rotation. The ratio of angular velocities $\mu=\hat{\omega}_i/\hat{\omega}_o$, where $\hat{\omega}_i$ ($\hat\omega_o$) is the inner (outer) cylinder angular velocity, can be written as:

\begin{equation}
 \mu = \displaystyle\frac{-\eta(1-\eta)+\eta R_\Omega}{(1-\eta) + \eta R_\Omega}
\end{equation}

\noindent where $\eta$ is the radius ratio $\eta=\hat{r}_i/\hat{r}_o$, and $\hat{r}_i$ ($\hat{r}_o$) is the radius of the inner (outer) cylinder.}

In this manuscript, we fix $Re_s=3.62\times10^4$ and vary $R_\Omega$ in the range $[-0.1,0.7]$, covering anti-cyclonic and cyclonic rotation. The domain is taken to be axially periodic, with a periodicity length $\hat{L}_z$, which can be expressed non-dimensionally as an aspect ratio $\Gamma=\hat{L}_z/\hat{d}$. This aspect ratio is fixed to $\Gamma=2.33$, enough to fit a single Taylor roll pair with a wavelength of $\lambda_{TR}=\Gamma=2.33$. We also fix the radius ratio to be $\eta=0.91$, resulting in a geometry which is usually considered a narrow gap. We also impose a rotational symmetry of order $n_{sym}=20$ on the system to reduce computational costs. For this radius ratio, this results in a streamwise periodicity length of around $2\pi$ half-gaps, large enough to obtain asymptotic torque and mean flow statistics \cite{ostilla2015effects}.

Spatial discretization is performed using a second-order energy-conserving centered finite difference scheme, while time is advanced using a low-storage third-order Runge-Kutta for the explicit terms and a second-order Crank-Nicholson scheme for the implicit treatment of the wall-normal viscous terms. More details of the algorithm can be found in \cite{verzicco1996finite, van2015pencil}. \textcolor{blue}{The code has been extensively validated for Taylor-Couette flow by matching torque and mean velocity statistics to experimental data \cite{ostilla2014exploring,ostilla2014boundary,ezeta2020double}.} The spatial resolution is based on \cite{sacco2019dynamics}, which uses $n_\theta\times n_r\times n_z = 384\times 512\times 768$ in the azimuthal, radial and axial directions respectively for a similar system. This corresponds to resolutions in wall-units of $\Delta z^+\approx 5$, $\Delta x^+ = r\Delta \theta^+ \approx 9$ and $\Delta r^+ \in (0.5, 5)$ for the most adverse conditions. We assess temporal convergence by measuring the difference in torque between both cylinders, and only when this difference is lower than $1\%$ the simulation is considered to have acceptable errors due to lack of temporal convergence.

For convenience, we define the non-dimensional wall distance coordinate as $\tilde{r}=(\hat{r}-\hat{r}_i)/\hat{d}$, and the non-dimensional axial coordinate $\tilde{z}=\hat{z}/\hat{d}$. We also note that results for rotating Waleffe flow are taken from \cite{farooq2020large} which uses the same numerical scheme. Full details on the choice of spatial resolutions and domain sizes can be found there.

\section{Results}

\subsection{Rolls and global statistics}

\begin{figure}[!h]
\centering
\includegraphics[width=0.48\textwidth]{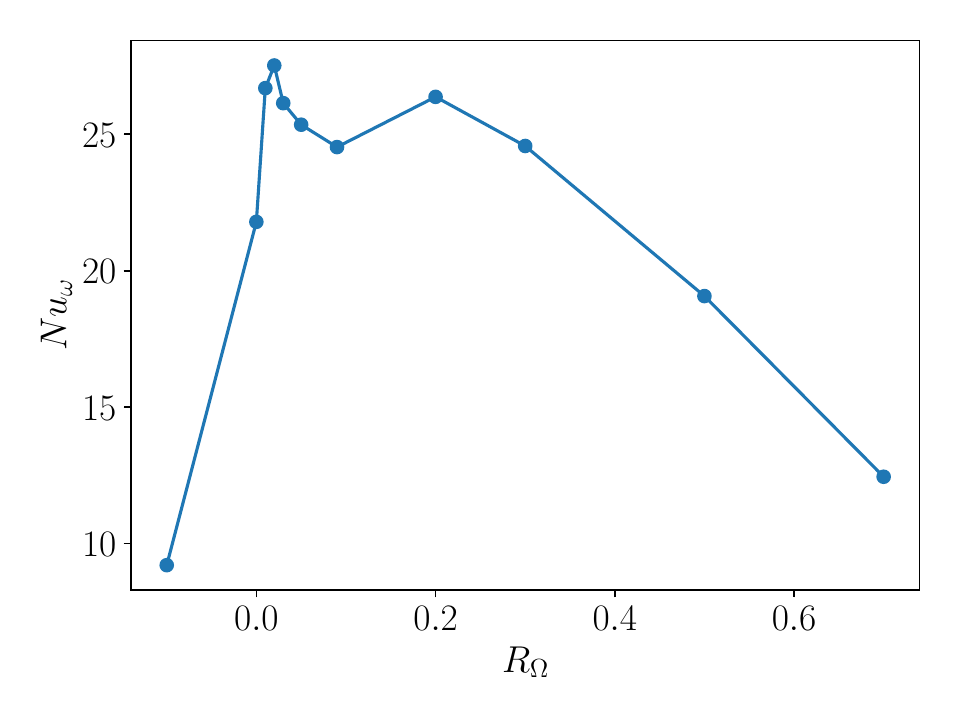}
\includegraphics[width=0.48\textwidth]{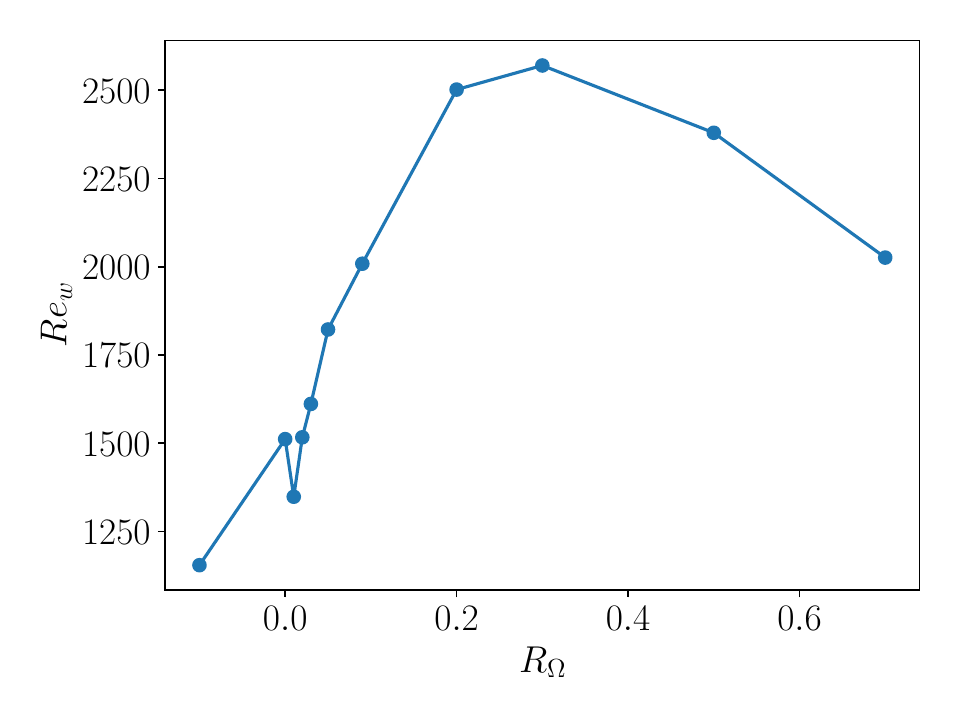}
\caption{Left panel: Torque non-dimensionalized as a Nusselt number $Nu_\omega$ against Coriolis parameter $R_\Omega$. Right: ``Wind'' Reynolds number $Re_w$ against Coriolis parameter $R_\Omega$.}
\label{fi:re_nu_vs_romega}
\end{figure}

To begin the analysis, we show the torque required to drive the cylinders non-dimensionalized as a Nusselt number $Nu_\omega=\hat{T}/\hat{T}_{pa}$, where $\hat{T}$ is the torque at the cylinders and $\hat{T}_{pa}$ is the torque for the purely azimuthal solution, as a function of the Coriolis parameter $R_\Omega$ in the left panel of figure \ref{fi:re_nu_vs_romega}. The behaviour of $Nu_\omega(R_\Omega)$ shows the ``double'' peak (or local maxima) structure previously observed in 
\cite{brauckmann2017marginally,ezeta2020double}, at similar Reynolds numbers. The first peak, called the ``narrow'' peak ($R_\Omega\approx 0.05$) is caused by boundary-layer shear instabilities that change the marginal stability of the system \cite{brauckmann2017marginally}. The second ``broad'' peak ($R_\Omega\approx 0.2$), is related to the strongest transport of momentum through Taylor roll-like structures \cite{brauckmann2017marginally}. As the Reynolds number further increases ($Re_s>10^5$), the two peaks become one \cite{ezeta2020double}. The boundary layers become unstable for all $R_\Omega$, and the remaining maximum in the $Nu_\omega$ curve is solely related to the largest transport of momentum through Taylor roll-like structures, the same mechanism that generated the broad peak \cite{brauckmann2017marginally,ezeta2020double}.

\textcolor{blue}{To quantify the turbulence levels, we define the  ``wind'' Reynolds number $Re_w$ \cite{eckhardt2007torque} as $Re_w = Re_s \sqrt{\langle u_r^2\rangle_{V,t} + \langle u_z^2 \rangle_{V,t}}$, where $\langle . \rangle_{V,t}$ denotes an averaging operation over the entire volume $V$ and time. We note that this definition of $Re_w$ is based on averaging the total energy of the radial and axial components of the instantaneous velocity, and is different from those using the variance of the radial velocity, as used in for example Ref.~\cite{huisman2012ultimate}. We choose this because due to the presence of large-scale structures which introduce axial inhomogeneities in the mean statistics, there is a degree of ambiguity when defining the variance, and this leads to different results depending on the exact method taken to calculate said variance \cite{ostilla2016near}. While this point is not important if one focuses on the $Re_w(Re_s)$ behaviour while keeping $R_\Omega$ constant, as in Ref.~\cite{huisman2012ultimate}, it complicates the analysis of $Re_w(R_\Omega)$ as the Taylor rolls (and hence the axial inhomogeneity which biases the results) disappears across varying $R_\Omega$. }

In the right panel of figure \ref{fi:re_nu_vs_romega}, we show $Re_w$ against $R_\Omega$. A large dip can be seen in the graph between $R_\Omega=0$ and $R_\Omega=0.01$ which we can associate to the stratification of velocity in the boundary layers: as rotation is slightly increased, some areas of the boundary layer become quiescent and the turbulence level drops substantially \cite{ezeta2020double}. Only as $R_\Omega$ is further increased do the turbulence levels recover, reaching a maximum at $R_\Omega=0.3$, before dropping for larger $R_\Omega$. We note that there is a small discrepancy between the $Re_w$ maximum and the $Nu_\omega$ broad peak, a phenomena we will rationalize later.

\begin{figure}[!h]
\centering
\includegraphics[trim={4.5cm 0 4.5cm 0cm}, clip, height=0.3\textwidth]{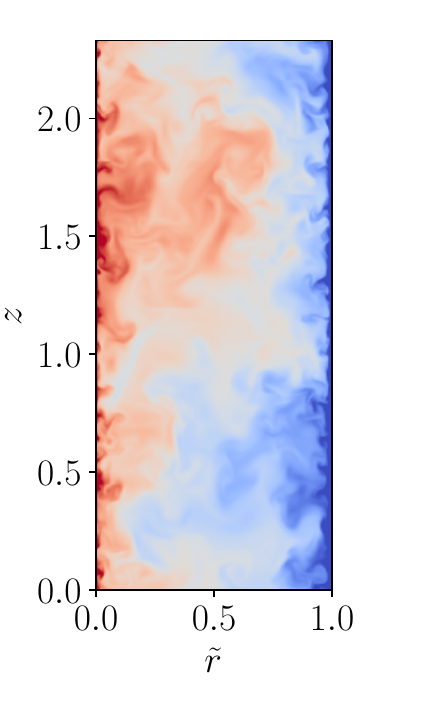}
\includegraphics[trim={4.5cm 0 4.5cm 0}, clip, height=0.3\textwidth]{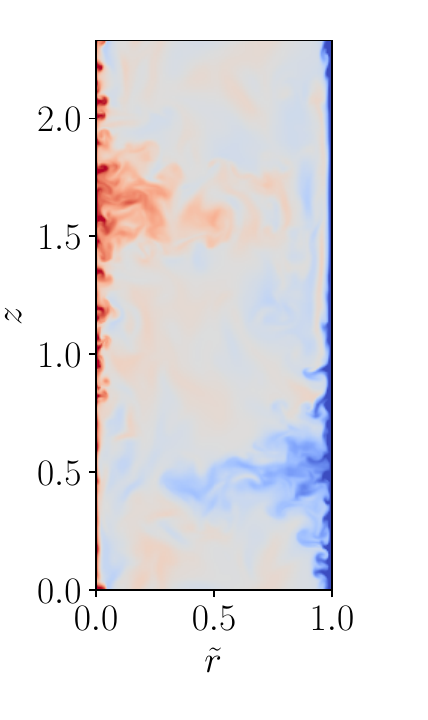}
\includegraphics[trim={4.5cm 0 4.5cm 0}, clip, height=0.3\textwidth]{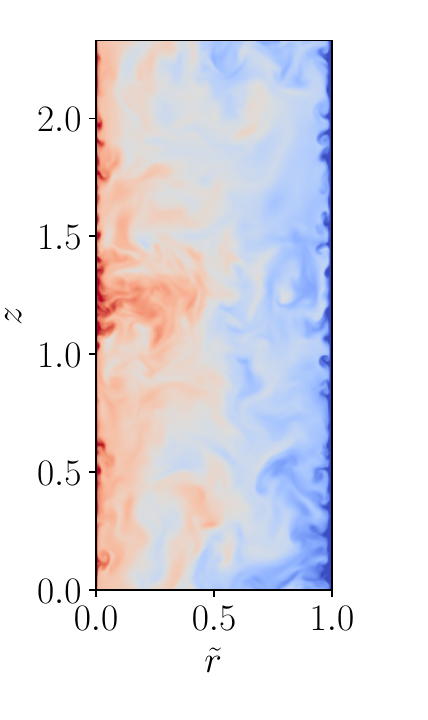}
\includegraphics[trim={7cm 0 0cm 0cm}, clip, height=0.3\textwidth]{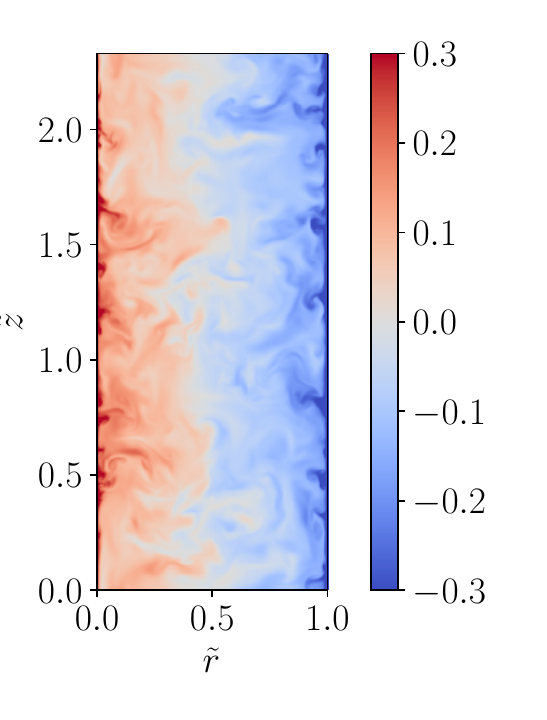}\\
\includegraphics[trim={4.5cm 0 4.5cm 0cm}, clip, height=0.3\textwidth]{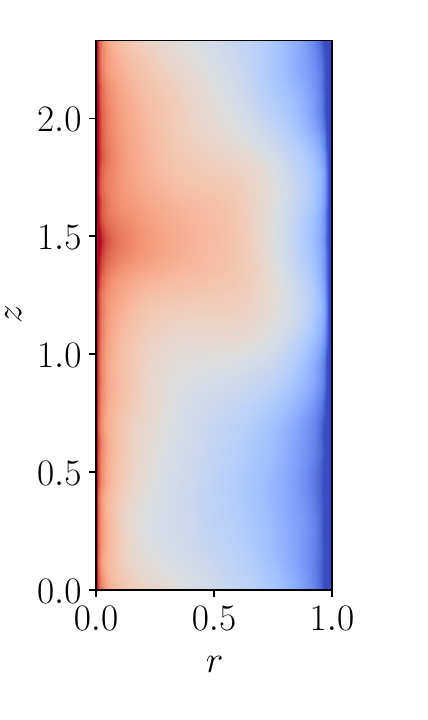}
\includegraphics[trim={4.5cm 0 4.5cm 0}, clip, height=0.3\textwidth]{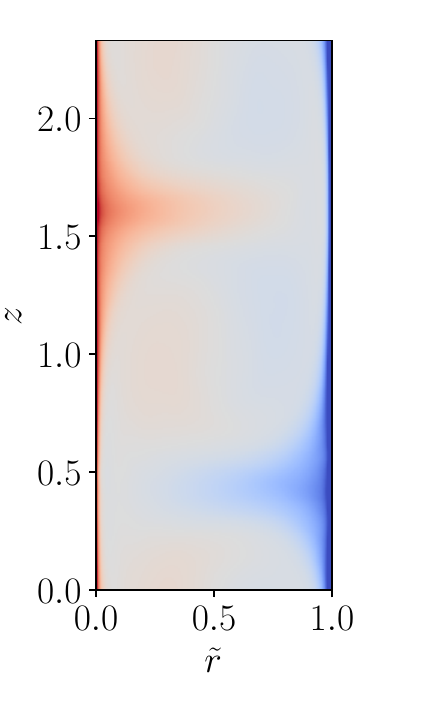}
\includegraphics[trim={4.5cm 0 4.5cm 0}, clip, height=0.3\textwidth]{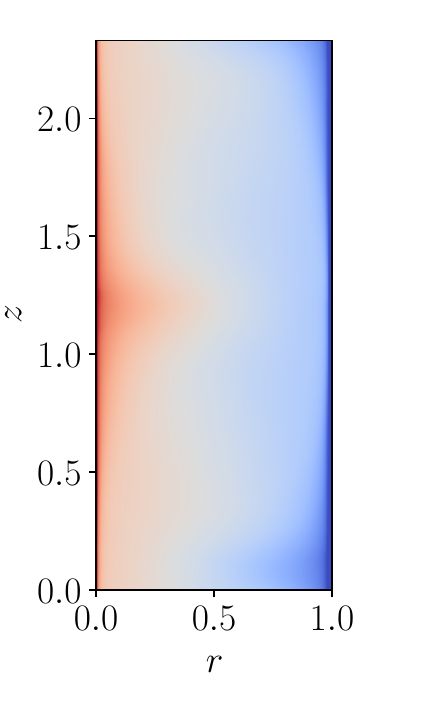}
\includegraphics[trim={7cm 0 0cm 0cm}, clip, height=0.3\textwidth]{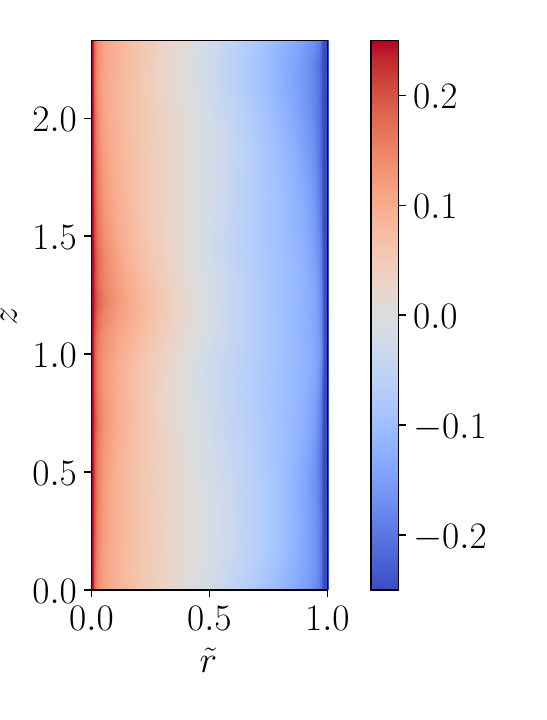}
\caption{Top row: pseudocolor visualization of the instantaneous azimuthal velocity $u_\theta$ at a constant azimuth cut for  $R_\Omega=0$, $0.09$, $0.2$, $0.3$ (left to right). Bottom row: pseudocolor visualization of the azimuthally and temporally averaged azimuthal velocity $\langle u_\theta \rangle_{\theta,t}$ for the same values of $R_\Omega$.}
\label{fi:thcut-q1inst}
\end{figure}

To analyze the formation of the Taylor rolls we will focus on the broad peak. We select four cases to study in more detail: $R_\Omega=0$, as the canonical non-rotating case, $R_\Omega=0.09$ where the roll is strongest and also corresponds to pure inner cylinder rotation in the laboratory frame, $R_\Omega=0.2$ where the roll is still present but less sharp and $Nu_\omega$ is maximum, and $R_\Omega=0.3$ when the roll has almost disappeared but $Re_w$ is largest. In figure \ref{fi:thcut-q1inst}, for these four cases we present both an azimuthal cut of the instantaneous azimuthal velocity in the top row, and the temporally- and azimuthally averaged azimuthal velocity $\langle u_\theta \rangle_{\theta,t}$ in the bottom row. As the Coriolis parameter is increased from zero, the fluctuations are organised and form the Taylor rolls, large-scale structures composed by fluctuations but visible on an average sense. This organization can be observed to be very sharp for $R_\Omega=0.09$, and more diffuse for $R_\Omega=0.2$. Further increasing the Coriolis force by setting $R_\Omega=0.3$ causes the organization to disappear. 

\begin{figure}[!h]
\centering
\includegraphics[trim={4.5cm 0 4.5cm 0cm}, clip, height=0.3\textwidth]{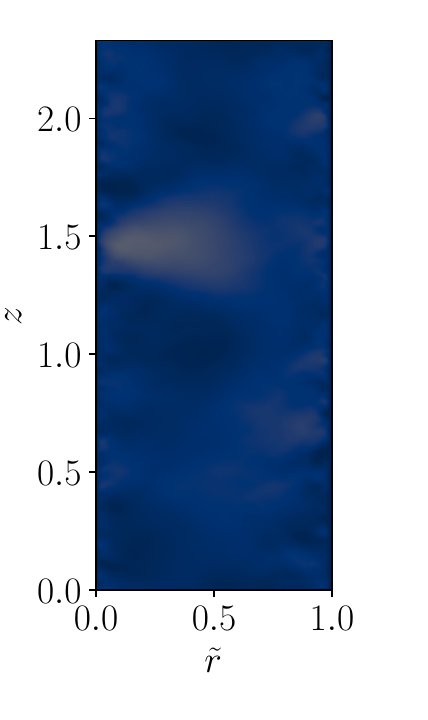}
\includegraphics[trim={4.5cm 0 4.5cm 0}, clip, height=0.3\textwidth]{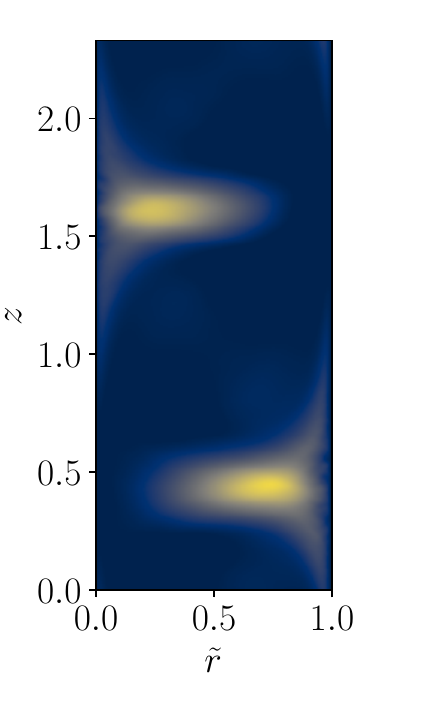}
\includegraphics[trim={4.5cm 0 4.5cm 0}, clip, height=0.3\textwidth]{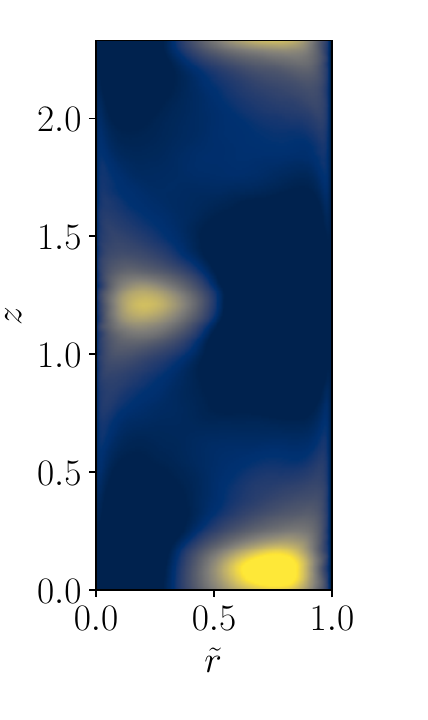}
\includegraphics[trim={7cm 0 0cm 0cm}, clip, height=0.3\textwidth]{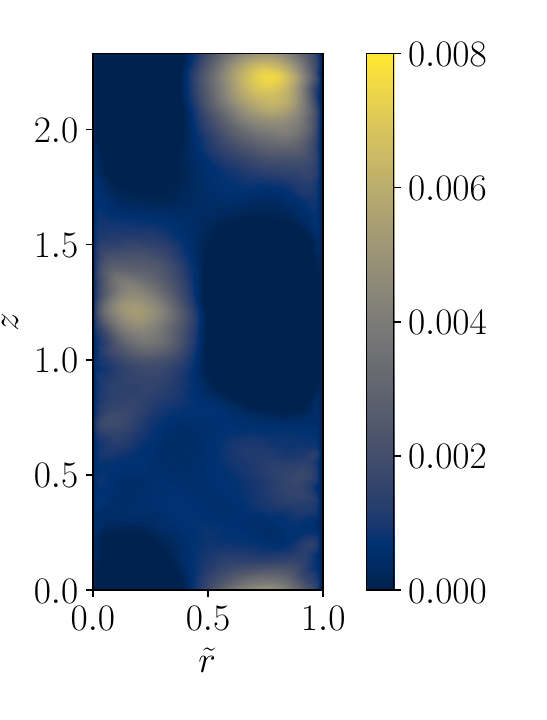}
\caption{Pseudocolor visualization of the azimuthally and temporally averaged convective transport $r^2 \langle u_r u_\theta \rangle_{\theta,t}$ for $R_\Omega=0$, $0.09$, $0.2$, $0.3$ (left to right).}
\label{fi:q12me}
\end{figure}

{\color{blue} To demonstrate how the Taylor rolls organize the torque transport across the cylinder gap, we can examine the convective transport of torque. We begin by noting that the angular velocity current $J^\omega$ \cite{eckhardt2007torque}, is defined as 

\begin{equation}
 J^\omega = r^2 \langle u_r u_\theta \rangle_{z,\theta,t} - Re_s^{-1} r^3 \partial_r \langle u_\theta/r \rangle{z,\theta,t}.
\end{equation}

\noindent This quantity is constant across the gap, equal to the torque at both cylinders, and composed of a convective term and a viscous term. We can identify the  term $r^2 (u_r u_\theta)$ with the convective transport, and average it temporally and azimuthally (but not axially) to show the effect of the rolls on transport. 

In figure \ref{fi:q12me} we show how $r^2\langle u_r u_\theta \rangle$  change as $R_\Omega$ is varied. While for $R_\Omega=0$ no apparent signature is present, showing that torque transport across the cylinders happens mainly through other mechanisms, for the other values of $R_\Omega$ the axial signature is more obvious.} Large values of this quantity coincide with the areas of the flow where fluctuations travel from inner to outer cylinder, or viceversa. This confirms what is reported in \cite{brauckmann2017marginally}: that when the rolls are present, they are responsible for a major part of the transport of angular velocity/momentum. \textcolor{blue}{We also note that due to the rotating frame formulation, the axially and temporally-averaged values of $r^2(u_\theta u_r)$ tend to always be positive. Calculating this term in a laboratory/inertial frame of reference will produce different results. }

\subsection{The effect of the two-way coupling}

{\color{blue} As mentioned in the introduction, for TCF, the Coriolis force induces a two-way coupling though a body force, while the body force in 2D3CRPCF and RBC is only one way. 

To clarify what we precisely mean by this, we turn to the Navier-Stokes equations. In TCF, the two-way coupling comes from the fact that the Coriolis term in the Navier-Stokes equation, when put on the right-hand-side, can be read as a body force with the following components:

\begin{equation}
    \textbf{f}_\Omega = -R_\Omega (\textbf{e}_z\times\textbf{u}) = -R_\Omega(u_r \textbf{e}_\theta - u_\theta \textbf{e}_r).
    \label{eq:tw-tc}
\end{equation}

\noindent Furthermore, this term not acts on two velocity components, but it acts in opposing directions: large values of $u_r$ tend to negatively force $u_\theta$, while large values of $u_\theta$ would (positively) force $u_r$. 

In comparison, only the second type of coupling is present in 2D3CRPCF and RBC: the wall-normal velocity has a positive force which is proportional to the streamwise velocity or temperature. That is, in an RBC system under the Bousinnesq approximation and where $z$ is the vertical coordinate, the non-dimensional body force due to buoyancy reads: 

\begin{equation}
    \textbf{f}_b = \theta \textbf{e}_z,
    \label{eq:ow-rb}
\end{equation}

\noindent where $\theta$ is the non-dimensional temperature. This force has a single component which is equivalent to the $R_\Omega u_\theta \textbf{e}_r$ term. No equivalent to the $-R_\Omega u_r \textbf{e}_\theta$ is present in this force.

This observation is particularly important because in TCF, the convective transport of the torque is given by the average of the product of $u_r$ and $u_\theta$ as discussed above (and similarly, in RBC the average convective transport of heat is given by the average of the product of $u_z$ and $\theta$). Hence, increased $R_\Omega$ cannot be identified as just an increased forcing, equivalent to simply increasing the temperature in an instance of RBC.} Instead, there must be some competing effects between both components of the Coriolis force. So then, the question arises: why do mild positive values of $R_\Omega$ increase transport and cause the velocity field to organize into rolls, while larger positive values cause a substantial drop in transport and this disappearance of the rolls' organization?

\begin{figure}[!h]
\centering
\includegraphics[width=0.32\textwidth]{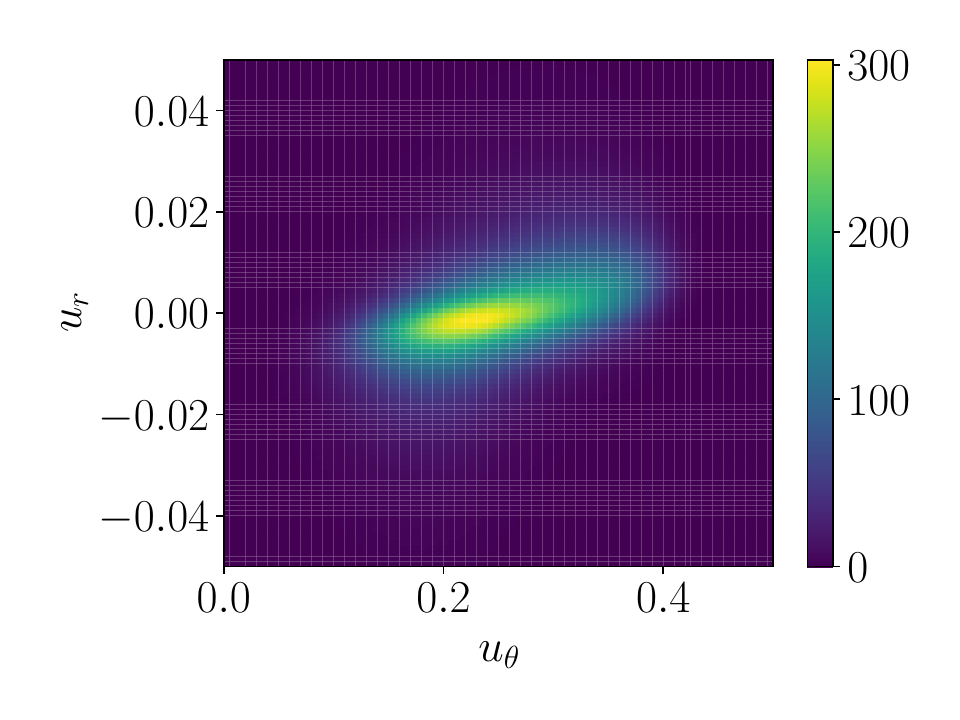}
\includegraphics[width=0.32\textwidth]{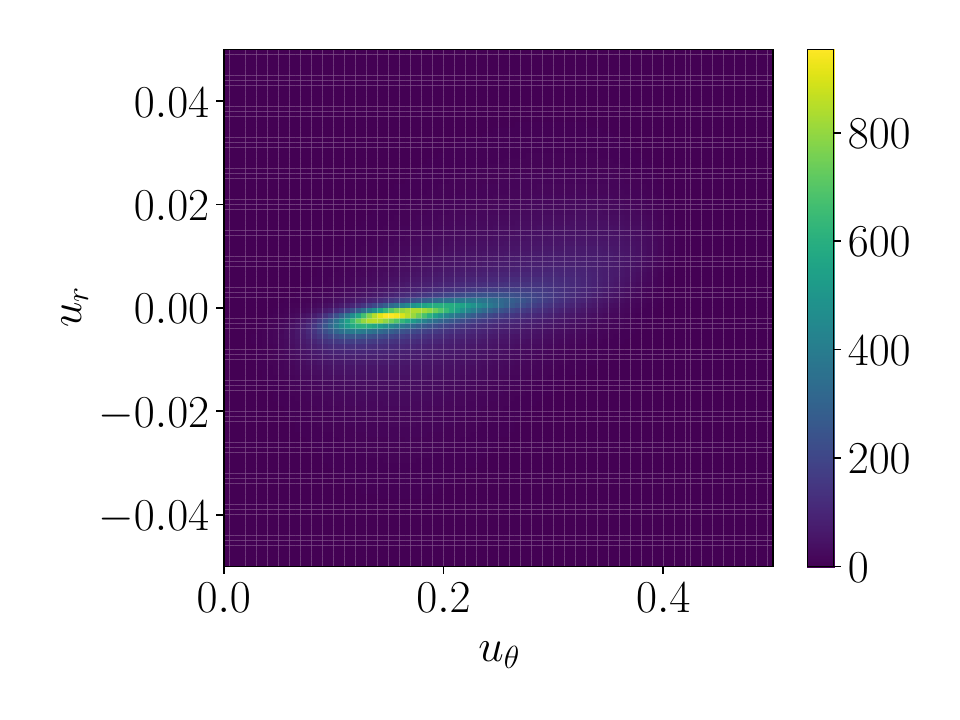}
\includegraphics[width=0.32\textwidth]{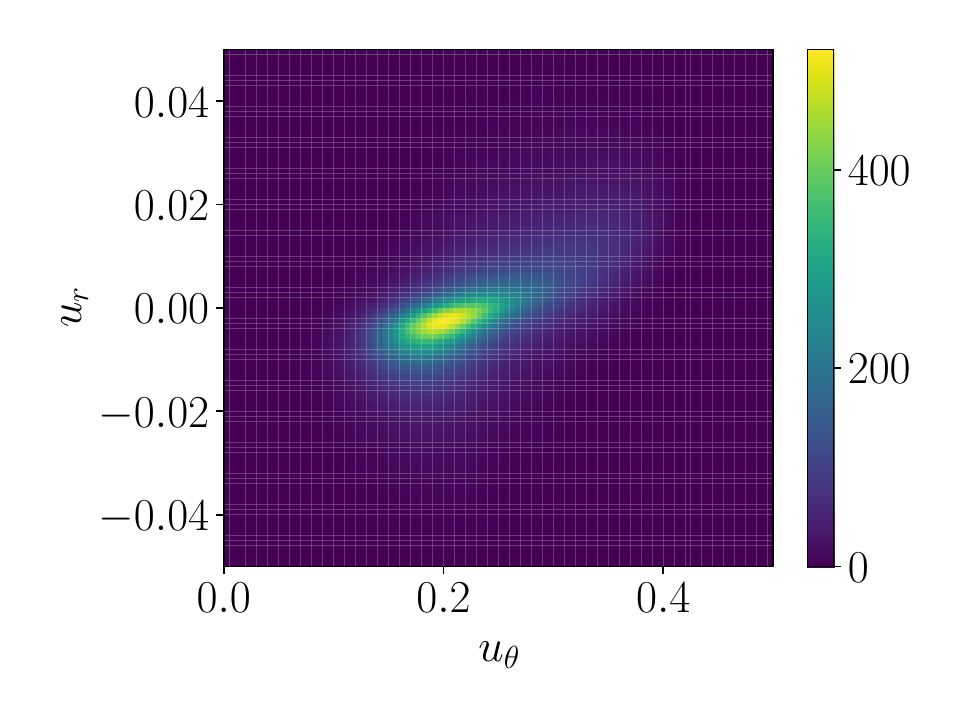}
\includegraphics[width=0.32\textwidth]{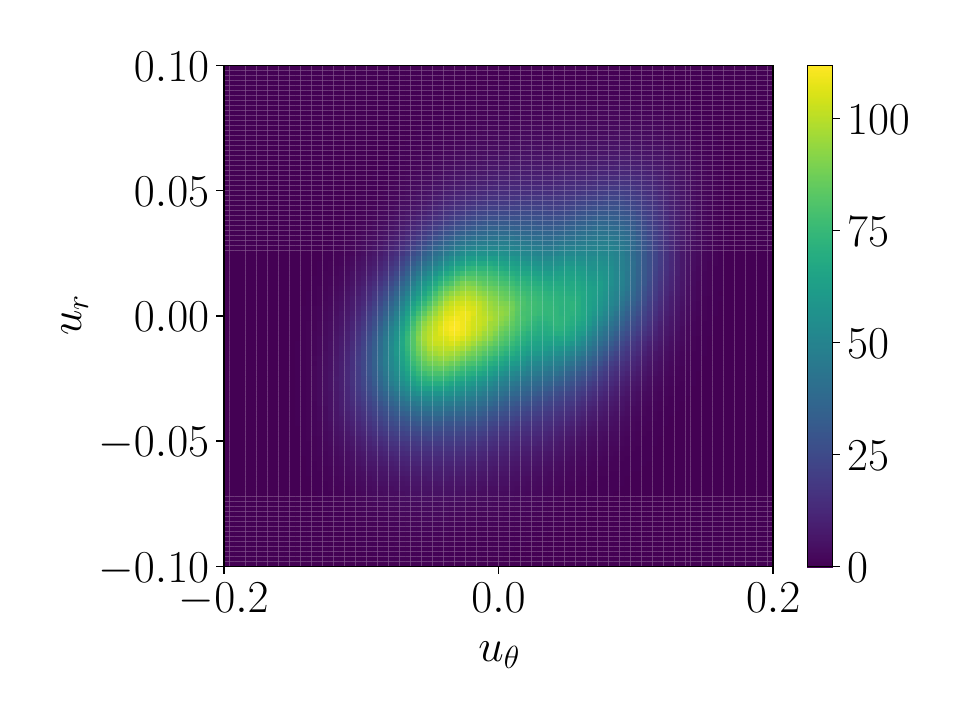}
\includegraphics[width=0.32\textwidth]{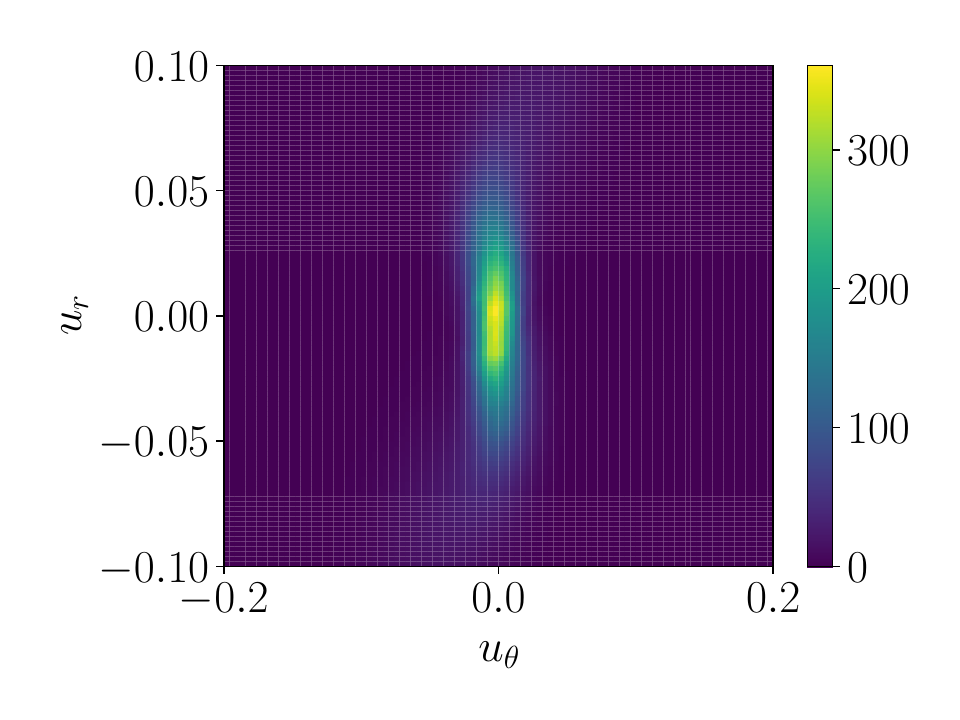}
\includegraphics[width=0.32\textwidth]{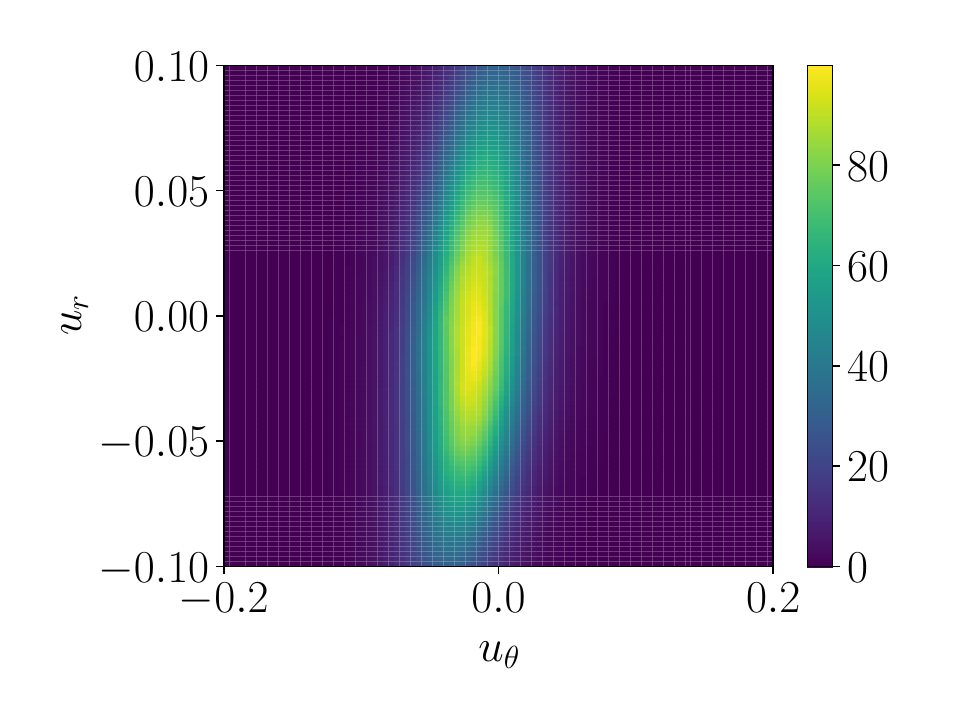}
\caption{Joint probability distribution function for radial and azimuthal velocities in the boundary layer (top row) and mid-gap (bottom row), for $R_\Omega=0$ (left), $R_\Omega=0.09$ (middle) and $R_\Omega=0.3$ (right).}
\label{fi:2dpdf-tc-bl}
\end{figure}

To explore this, we begin by looking at a point where we expect the two-way coupling to be at its weakest: inside the boundary layer. This is because we expect the magnitudes of $u_r$ to be much smaller than those of $u_\theta$, and hence the coupling to be much stronger in the radial velocity, being effectively one-way and similar to RBC. In the top row of figure \ref{fi:2dpdf-tc-bl}, we show the joint probability distribution for $u_r$ and $u_\theta$, inside the boundary layer at $\tilde{r}=0.013$ (corresponding to $r^+\approx 13$ in viscous units) for three values of $R_\Omega$. While the $R_\Omega=0$ and $0.2$ distributions are more spread out, the $R_\Omega=0.09$ distribution is extremely concentrated due to the presence of quiescent regions in large parts of the boundary layer. A bias, where high velocity regions are correlated with outflow velocities ($u_r>0$) can be appreciated even for the non-rotating case, due to the natural turbulence of a shear boundary layer. As the rotation parameter increases, the Coriolis body force induces a stronger coupling between radial and azimuthal velocity. This is reflected by a larger ``tail'' in the upper quadrant for large values of $u_\theta$ and positive values of $u_r$. 

As we turn to the bulk, the radial and azimuthal velocities become of approximately the same order. Hence, if our earlier observations on the contradictory nature of the two-coupling are correct, we would expect the bias to disappear \textcolor{blue}{as one term of the body force would increase a positive correlation and the other would tend to increase a negative correlation}. Indeed, this is what is seen in the bottom row of figure \ref{fi:2dpdf-tc-bl}, which shows the same joint probability distribution functions at $\tilde{r}=0.5$, i.e~the mid-gap. The shape of the joint probability distribution is now very different from what is seen for the non-rotating case ($R_\Omega=0$). In addition, when $R_\Omega\neq 0$, the range of $u_\theta$ is reduced while the values of $u_r$ are much larger, showing that rotation heavily amplifies the ``wind'' levels as was shown in figure \ref{fi:re_nu_vs_romega}.

\begin{figure}[!h]
\centering
\includegraphics[width=0.49\textwidth]{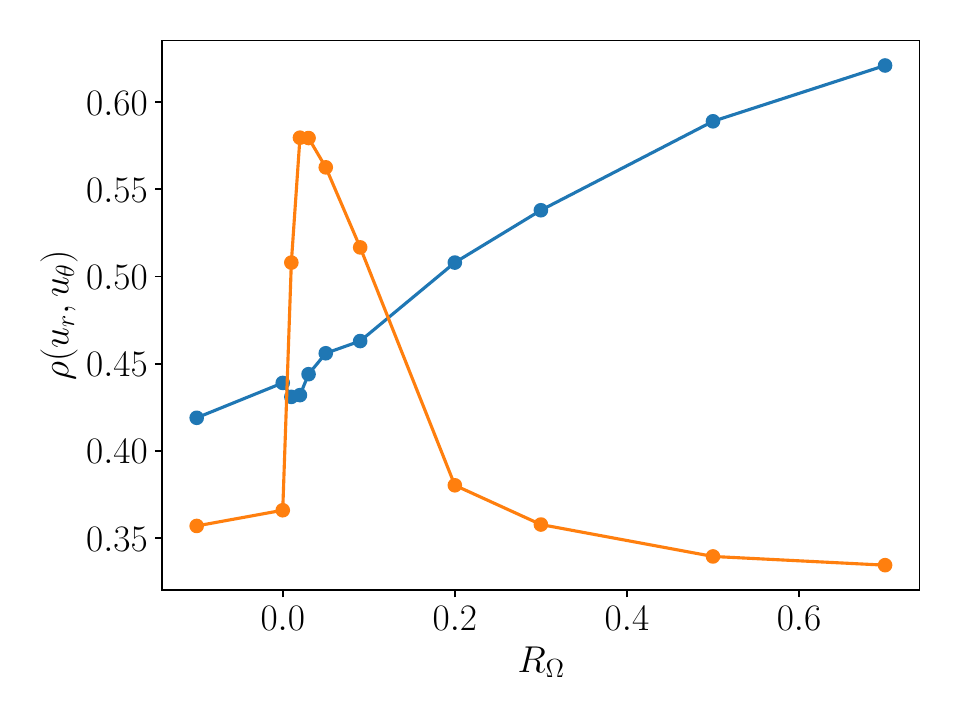}
\includegraphics[width=0.49\textwidth]{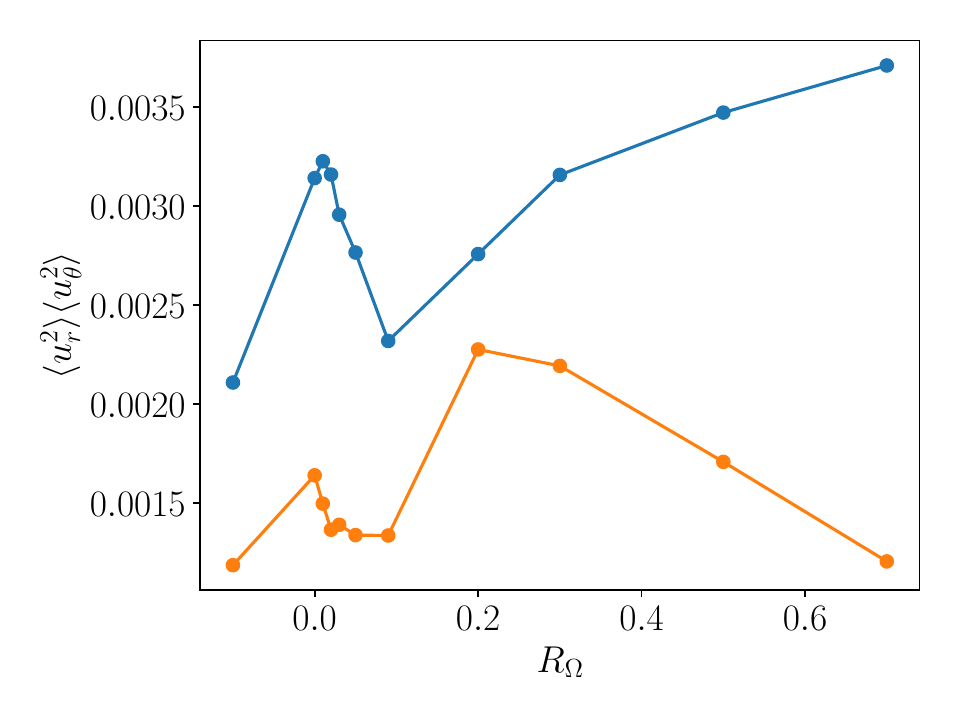}
\caption{Left panel: Pearson correlation coefficient between the radial and azimuthal velocities as a function of $R_\Omega$. Right panel: Product of the second moment of the azimuthal and radial velocities as a function of $R_\Omega$. Symbols: boundary layer ($\tilde{r}=0.013$) is the blue line and the mid-gap ($\tilde{r}=0.5$) is the orange line. }
\label{fi:ros_vs_correlation}
\end{figure}

To further quantify these effects, we calculate the Pearson correlation coefficient $\rho$ between radial and azimuthal velocities as a function of $R_\Omega$ in the boundary layer ($r^+\approx 13$) and bulk ($\tilde{r}=0.5$), and show this in the left panel of figure \ref{fi:ros_vs_correlation}. We can see an almost monotonic increase in the correlation coefficient in the boundary layer. This confirms our theory that $R_\Omega$ induces a strong coupling between radial and azimuthal velocities in the boundary layer. For the bulk region, the correlation coefficient is lower, but still positive, reflecting the transport of angular velocity (momentum) by fluctuations. More notably, correlation is not a strong function of $R_\Omega$ except in the region $0<R_\Omega<0.2$, which coincides with the presence of rolls (figure \ref{fi:thcut-q1inst}). Therefore, the rolls are related to strong correlations between azimuthal and radial velocities, as could be expected from visual inspection of figures \ref{fi:thcut-q1inst}-\ref{fi:q12me}. This is similar to what is observed for example in three-dimensional RBC where the temperature and vertical velocity fields look very similar in the mid-gap when large-scale structures are present \cite{krug2020coherence}.

The left panel of figure \ref{fi:ros_vs_correlation} raises one question. At the mid-gap, where transport is almost exclusively convective, the correlation coefficient is lower for $R_\Omega=0.2$ than $R_\Omega=0.1$. However, the total convective transport, measured as $Nu_\omega$ is higher for $R_\Omega=0.2$. To answer this, we show the product of the second moments of the radial and azimuthal velocities in the right panel of figure \ref{fi:ros_vs_correlation}. Here, we can clearly see that even if the correlation coefficients can be higher, overall, the total energy of the flow is lower, so there is less convective transport. This increase in total flow energy is consistent with the higher values of $Re_w$ observed earlier in figure \ref{fi:re_nu_vs_romega}. We conclude that optimum transport comes not only from the highest correlation between radial and azimuthal velocity, but is also related to high turbulence levels.

As correlation does not imply causation, it is not clear whether the rolls themselves are caused by carefully selecting an $R_\Omega$ that would yield a specific correlation between velocities sufficient to induce rolls in the boundary layer but not large enough to destroy them in the bulk, or if the mid-gap velocity correlations are simply a reflection of the presence of rolls. To further explore this, we turn to rotating Waleffe flow, another system where rolls are present, but where the velocity statistics look very different.

\subsection{Removing Boundary Layers: Waleffe Flow}

Rotating Waleffe flow (RWF, \textcolor{blue}{sketch in Fig.~\ref{fi:sch}) is the flow between two stress-free plates driven by a sinusoidal force \cite{waleffe1997self}. It is a canonical shear flow, generally similar to TCF/RPCF except for two aspects. The first, is that the no-slip condition is removed, and replaced by a no-penetration condition, so the nature of the boundary layers is changed, and shear is transported throughout the flow only (and not to the walls).} The second, is that the shear/torque transport is not constant, but is a function of plate distance, reaching a maximum at the center and zero at the plates, as the shear must balance the sinusoidal forcing (c.f.~Ref.~\cite{farooq2020large} for a longer discussion). In this second aspect RWF could also be thought of an analog to pressure-driven channel flow, which has a non-constant transport of shear.

RWF is of interest in this analysis as it has been shown to have large-scale stream-wise invariant structures, very reminiscent of Taylor rolls, when a mild anti-cyclonic rotation is present \cite{farooq2020large}. The structures are slightly different: possessing a ``full'' vorticity core with a maximum vorticity at the center of the structure rather than at the edges (cf. Fig.~14 in \cite{farooq2020large}), but they serve a similar purpose, transporting shear, and can be visually seen to be present when there is a strong coupling between wall-normal and streamwise velocities (cf. Fig.~8 in \cite{farooq2020large}).

\begin{figure}[!h]
\centering
\includegraphics[width=0.32\textwidth]{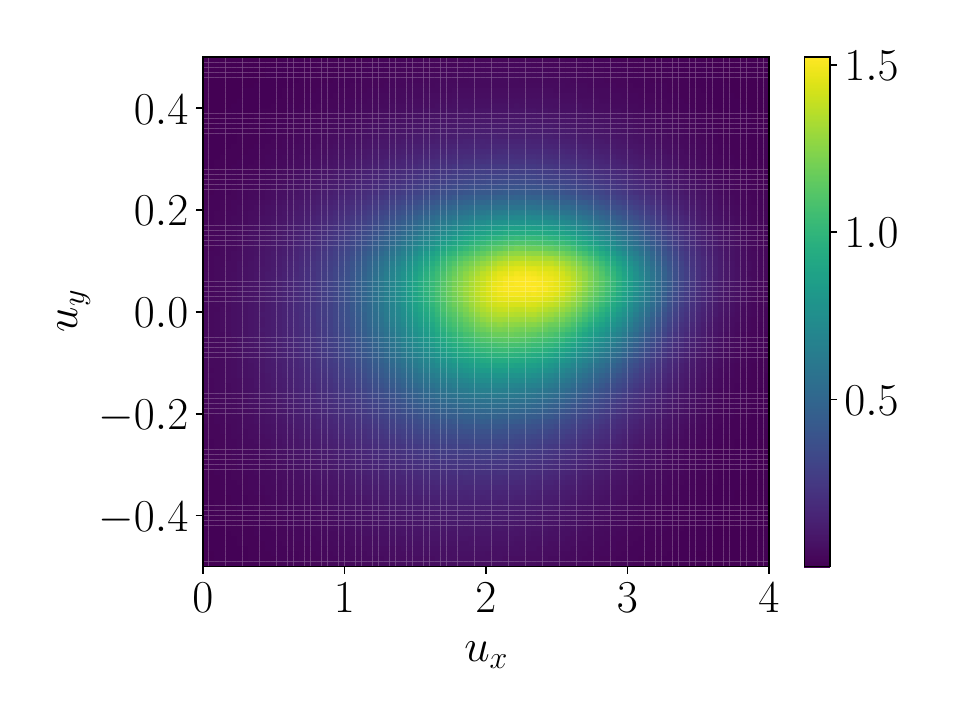}
\includegraphics[width=0.32\textwidth]{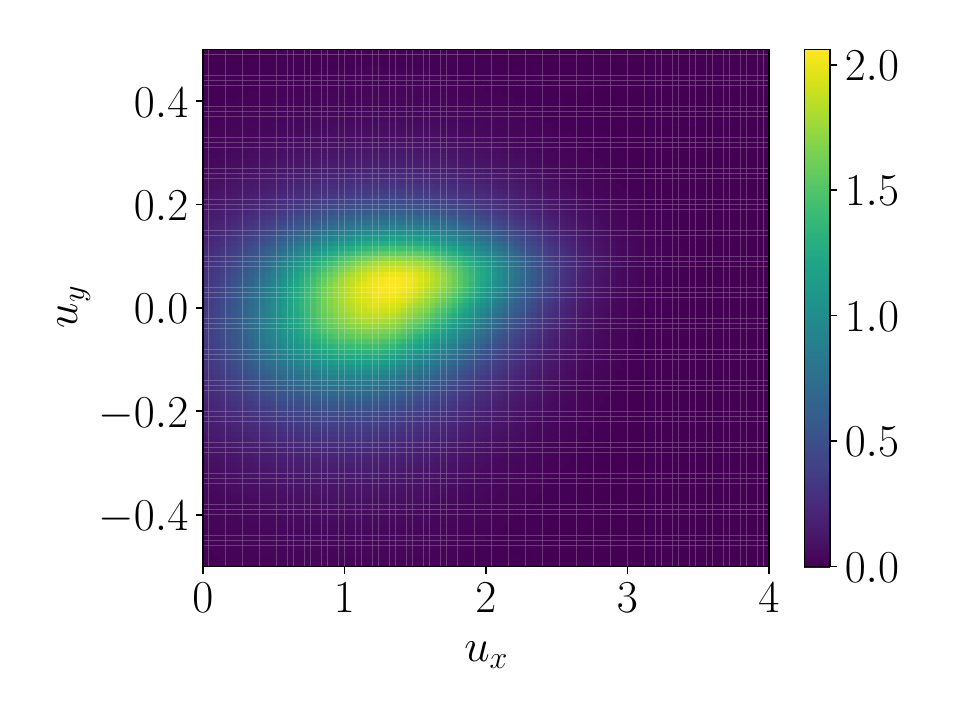}
\includegraphics[width=0.32\textwidth]{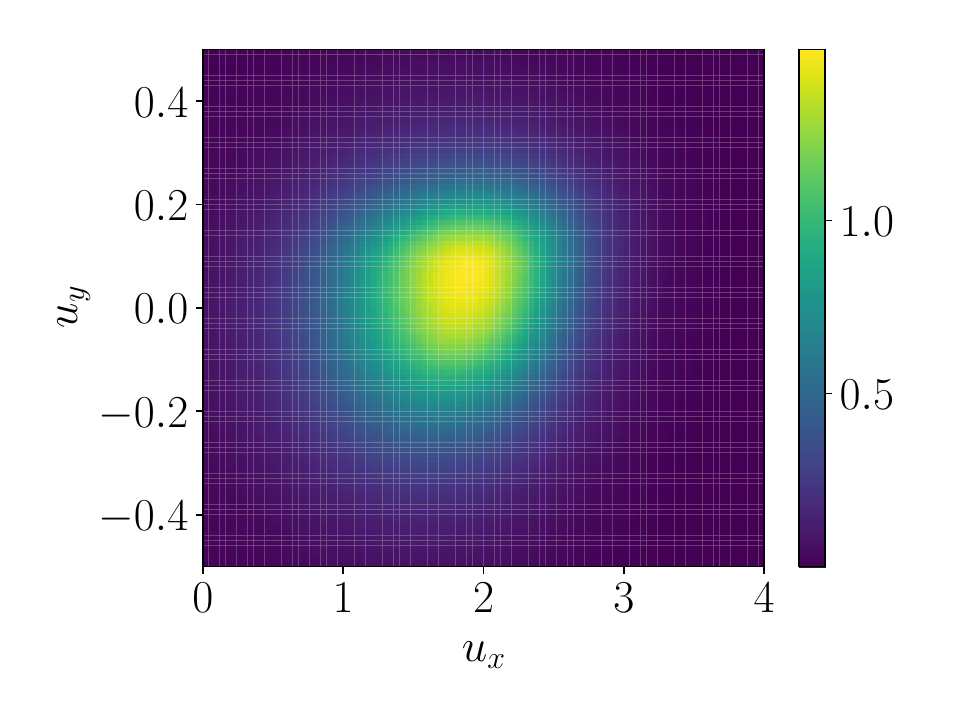}
\includegraphics[width=0.32\textwidth]{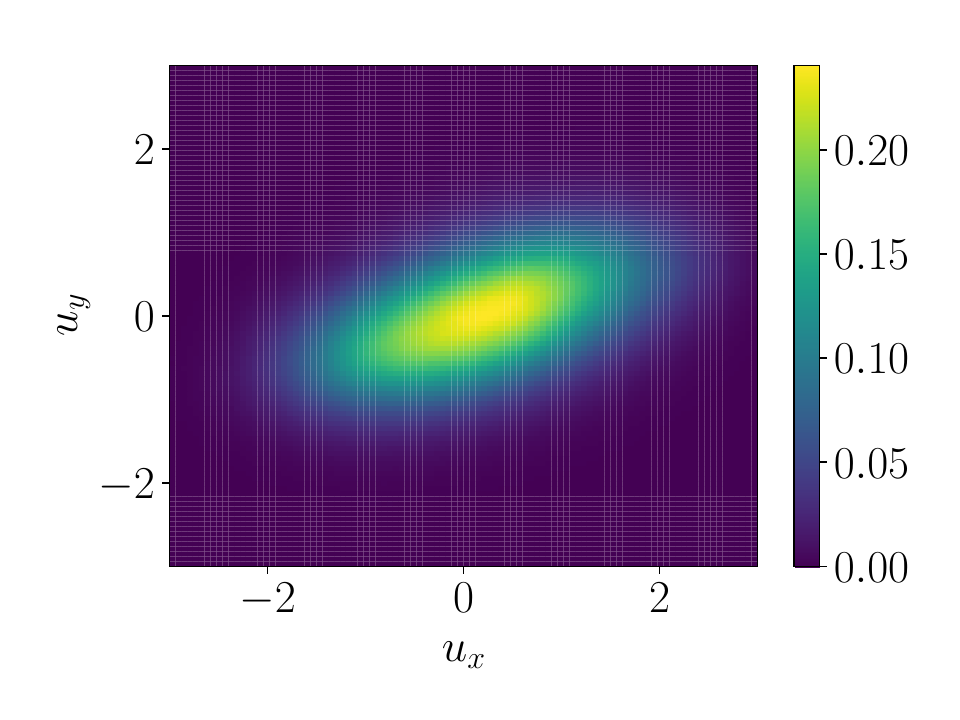}
\includegraphics[width=0.32\textwidth]{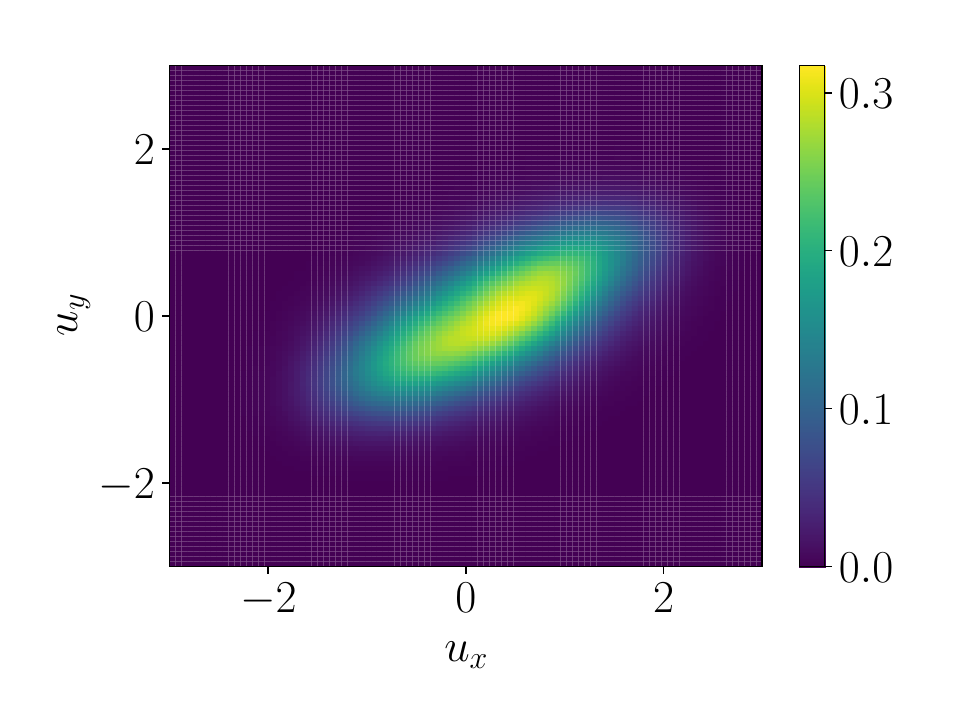}
\includegraphics[width=0.32\textwidth]{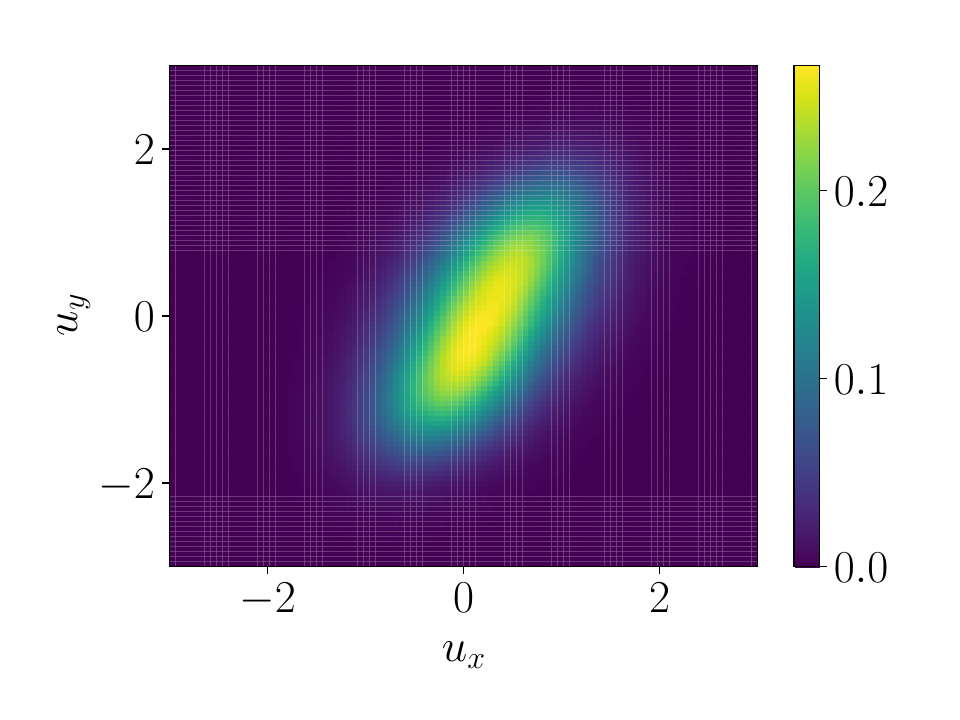}
\caption{Joint probability distribution function for wall-normal and streamwise velocities for RWF near the bottom stress-free plate (top row) and at the mid-gap (bottom row), for $R_\Omega=0$ (left), $0.63$ (middle) and $2.21$ (right).}
\label{fi:2dpdf-wf}
\end{figure}

In this subsection, we conduct a similar statistical analysis to the one conducted above for TCF, to check whether the two-way coupling framework helps us elucidate the origin of the rolls. In figure \ref{fi:2dpdf-wf} we show the joint probability distribution of the wall-normal velocity ($u_y$, analogous to $u_r$ in TCF) and the streamwise velocity ($u_x$, analogous to $u_\theta$ in TCF) for three rotation parameters near the stress-free plate ($y=0.013$) and at the mid-gap ($y=0.5$). The values of $R_\Omega$ are chosen to be the non-rotating case $R_\Omega=0$, the case where the large-scale rolls are strongest $R_\Omega=0.63$, and a case where the rolls are washed away due to the rotation being too large $R_\Omega=2.21$, similar to what was chosen above for TCF.

\begin{figure}[!h]
\centering
\includegraphics[width=0.49\textwidth]{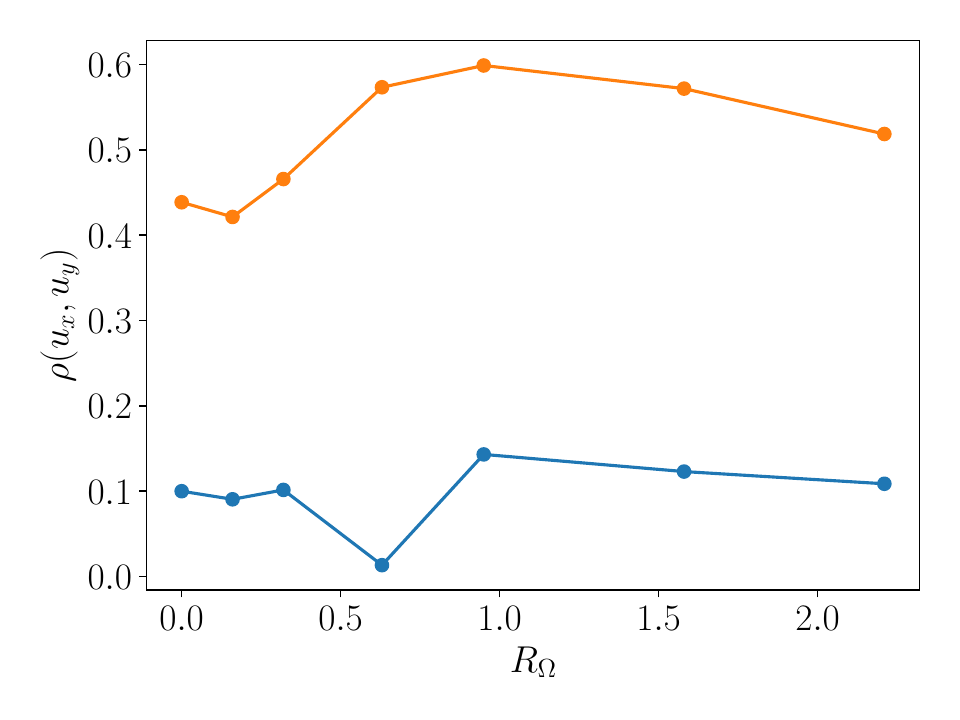}
\caption{Pearson correlation coefficient between the wall-normal and streamwise velocities for RWF as a function of $R_\Omega$. Symbols: blue line is the boundary layer location $y=0.013$ and orange line is the mid-gap location $y=0.5$.}
\label{fi:corr-wf}
\end{figure}

{ \color{blue} Interestingly, the trend seen for RWF is completely reversed. There is very little apparent coupling in the boundary layer, despite the non-penetration condition $u_y=0$ at the wall, which (similar to TCF) should cause relatively small wall-normal velocities in this region and the Coriolis body force to only act in one direction. Furthermore, the coupling is strong in the mid-gap, where the velocities have a much larger overall magnitude, and we would expect the body force to act in both directions.

We can quantify these effects by showing the Pearson correlation coefficient in figure \ref{fi:corr-wf}. The trends shown in this graph are very different from those previously shown in figure \ref{fi:ros_vs_correlation}. The correlation is highest in the bulk, and for all values of $R_\Omega>0.6$ reaches values of above 0.5 regardless of the presence of rolls. These values were only reached in TCF when rolls were present. Furthermore, the correlation is extremely small in the boundary layers, hovering around $0.1$, and is not strongly dependent on $R_\Omega$ as in the case of TCF, showing that the coupling framework does not make accurate predictions for this flow.}

{\color{blue} This is indeed a major difference between RWF and TCF:  in RWF $u_x$ and $u_y$ inside the boundary layer decouple, and this is reflected as low correlation coefficients. Once one looks at the conservation properties of the flow, this decoupling becomes less surprising: in RWF, convective transport of shear happens through the average of the product of $u_x$ and $u_y$ velocities, i.e.~$\langle u_x u_y \rangle$. This term is at a minimum at the boundaries and maximum near the mid-gap \cite{farooq2020large}. Therefore, we could trivially expect correlations to be higher where the transport is higher.}

However, this goes against our intuitions developed earlier using the two-way coupling framework, which should translate unmodified to RWF as the Navier-Stokes equations are the same (except for the constant shear forcing). In RWF, the wall-normal velocity $u_y$ should be very small, while $u_x$ is comparatively less constrained due to the absence of a no-slip condition. So we would expect the coupling to be close to one-way in this region, but this is not the case. 

This discrepancy may be caused by the generation of turbulence in different flow locations. Due to the presence of a no-slip condition, in TCF perturbations are formed very close to the wall, and travel from the boundary to the bulk. The direction of transport in RWF may be different, even if the original model by Waleffe  showed no substantial difference between no-slip and free-slip walls for the stability of the base profile and the formation of streaks and rolls \cite{waleffe1997self}. However, at high $Re$ different mechanisms could be at play. A full explanation would require a detailed study of the way perturbations form in RWF and this generation is different from the process in TCF, and this falls outside the scope of the current study. 

Turning to the correlation coefficient at the mid-gap, figure \ref{fi:corr-wf} shows that high correlations between radial/wall-normal velocity and azimuthal/streamwise velocity can be present in the absence of rolls. For example, $R_\Omega=2.21$ has a correlation coefficient of $\approx 0.50$ yet large-scale organization is completely absent. This shows that the two-way coupling framework is not useful for understanding roll formation in RWF, and suggests that it is not a productive framework to use for TCF either. These results highlight that an analysis of the roll formation must take into account the precise manner the fluctuations (or thermal plumes in the case of RBC) are formed and organized. The two-way coupling framework, while giving some insights on a major difference between 3D TCF and the other flows where rolls do not vanish is of limited applicability. 

\subsection{Additional remarks on the origins of the structures}

In the previous sections, we have attempted to find a relationship between the large-scale structures found in convective flows, such as RBC, which are related to the presence of hot rising and cold falling plumes, and the large-scale structures in TCF, which are related to the ejection of high-momentum structures from the boundary layers of the cylinders. From our analysis, it is has become more clear that the RBC-TCF analogy is far from perfect, and that in the turbulent regime the rough equivalence between large-scale thermal convective structures and Taylor rolls breaks down. Due to the presence of a two-way coupling, the solid body force that causes the pinned large-scale structures to arise is not easily assimilated to the \textcolor{blue}{buoyancy force present in RBC (compare Eqs.~\ref{eq:tw-tc} and \ref{eq:ow-rb})}. Instead, it leads to further questions.

A possible answer to the mystery is that the large-scale rolls have their origin in other families of structures found in shear flows. Non-rotating Plane Couette flow is known to have large-scale structures with lengths that greatly exceed the gap width \cite{tsukahara2006dns}. Through the use of auto-correlations, large-scale structures can also be found in non-rotating Waleffe flow simulations, even if they are not as long as those seen for PCF \cite{farooq2020large}. These structures are harder to capture than streamwise invariant, spanwise fixed rolls, but can be detected through the use of conditional averages \cite{avsarkisov2014turbulent}. It very well may be that in the turbulent regime, what we know as Taylor roll is simply this large-scale shear structure which changes shape once mild anti-cyclonic rotation is added, before vanishing totally once rotation becomes sufficiently large. 

This is further supported by an interesting property of the RWF structures: they are eigenmodes of the Laplacian operator \cite{farooq2020large}. This is a property they share with other turbulent secondary flows such as those in square ducts \cite{pirozzoli2018turbulence}, and could be a further indication that the turbulent Taylor rolls are part of a larger family of structures which are related to Reynolds stress imbalances and not to centrifugal instabilities. Further research along these lines will surely help elucidate the origin of the rolls.

\section{Summary and Outlook}

In the 100 years since Taylor's seminal paper was written, the amount of articles written on the flow between two cylinders has exploded. A Google Scholar search throws $\sim 48400$ references for Taylor-Couette at the time of writing, July 2022. The large-scale structures which now take the name of Taylor rolls have been better understood, yet their origin in the turbulent case remains a mystery. 

\textcolor{blue}{In this manuscript, we have attempted to understand their origin by rationalizing solid-body rotation as a body force that induces a two-way coupling in Taylor-Couette flow, and distinguishing this two-way coupling from the one-way couplings caused by buoyancy forces in convective flows such as Rayleigh-B\'enard.} We demonstrated how in the boundary layers of TCF, this coupling is essentially one way, and this causes a strong correlation between azimuthal and wall-normal velocities, but that this correlation disappears in the bulk in the absence of Taylor rolls. By comparing TCF to rotating Waleffe flow, we showed that a strong correlation between velocities is not related to the presence of large-scale structures, and we must search elsewhere for the origins of the rolls.

In order to properly address this question, we need a way to better track the formation of these structures across changes in $R_\Omega$, from the non-rotating case to the case where the rolls appear strongest. Developing filters that can adequately isolate the structures has been a recent area of interest for the authors \cite{sacco2020dynamic,nguyen2020visualization}. It is easy to ``see'' a Taylor roll, as it is streamwise invariant. Yet structures in non-rotating Plane Couette flow are much harder to visualize (c.f.~Ref.~\cite{avsarkisov2014turbulent} for an attempt). Tracking structures across shape changes is the next natural step if we are to understand their origins.

\enlargethispage{20pt}



\aucontribute{ KA and ROM conceived of and designed the study. VJ carried out the simulations. VJ and ROM carried out the data analysis. All authors drafted, read and approved the manuscript.}

\competing{We report no conflict of interest.}

\funding{We acknowledge funding from the National Science Foundation through grant NSF-CBET-1934121.}

\ack{We acknowledge the Research Computing Data Core, RCDC, at the University of Houston for providing us with computational resources and technical support.}

\bibliographystyle{RS}
\bibliography{thebiblio}







\end{document}